\documentclass[runningheads]{llncs}
\usepackage[T1]{fontenc}
\usepackage{graphicx}
\usepackage{float}
\usepackage{mathtools}
\usepackage{amsmath}
\usepackage{amssymb}
\usepackage{color}
\usepackage{pythonhighlight}
\begin{document}
\title{Quantum Transfer Learning for MNIST Classification Using a Hybrid Quantum-Classical Approach}
\titlerunning{Quantum TL for MNIST Classification Using a Hybrid QC Approach}
\author{Soumyadip Sarkar}
\authorrunning{S. Sarkar}
\institute{}
\maketitle              % typeset the header of the contribution

\begin{abstract}
We implement a hybrid quantum-classical model for image classification that compresses MNIST digit images into a low-dimensional feature space and then maps these features onto a 5-qubit quantum state. First, an autoencoder compresses each $28\times28$ image (784 pixels) into a 64-dimensional latent vector, preserving salient features of the digit with minimal reconstruction error. We further reduce the latent representation to 5 principal components using Principal Component Analysis (PCA), to match the 5 available qubits. These 5 features are encoded as rotation angles in a quantum circuit with 5 qubits. The quantum feature map applies single-qubit rotations ($R_y$ gates) proportional to the feature values, followed by a Hadamard gate and a cascade of entangling CNOT gates to produce a non-product entangled state. Measuring the 5-qubit state yields a 32-dimensional probability distribution over basis outcomes, which serves as a quantum-enhanced feature vector for classification. A classical neural network with a softmax output is then trained on these 32-dimensional quantum feature vectors to predict the digit class. We evaluate the hybrid model on the MNIST dataset and compare it to a purely classical baseline that uses the 64-dimensional autoencoder latent features for classification. The results show that the hybrid model can successfully classify digits, demonstrating the feasibility of integrating quantum computing in the classification pipeline, although its accuracy (about 75\% on test data) currently falls below the classical baseline (about 98\% on the same compressed data). We provide a rigorous analysis of each component of the system autoencoder compression quality, quantum state encoding and measurement, and classifier performance and discuss the insights gained from the quantum measurement distributions. This work highlights how quantum feature mappings can transform classical data and provides a foundation for more advanced hybrid quantum-classical learning systems, while noting that further improvements and larger quantum circuits would be needed to surpass classical performance.

\end{abstract}

\section{Introduction}
Machine learning has achieved remarkable success in image recognition, exemplified by the MNIST handwritten digit classification problem which can reach high accuracy with deep neural networks~\cite{lecun1998,goodfellow2016}. However, training large models on high-dimensional data can be computationally intensive and prone to overfitting, a manifestation of the \emph{curse of dimensionality}~\cite{hinton2006}. Each MNIST image is a $28\times28$ grid of pixels (784-dimensional vector), and directly training on these raw features requires substantial computational resources and careful regularization. One effective strategy to tackle high dimensionality is to perform dimensionality reduction while preserving the important structure in the data. \emph{Autoencoders}, a class of neural networks for unsupervised representation learning, have proven capable of compressing data into low-dimensional latent codes that retain most of the variance or salient features of the input~\cite{baldi2012,vincent2008}. By training an autoencoder to minimize the reconstruction error of the input, one can obtain a compact feature vector (latent representation) that serves as a distilled version of the original data. In contrast to linear techniques like Principal Component Analysis (PCA), autoencoders can learn complex non-linear feature mappings and often achieve better compression for downstream tasks~\cite{hinton2006,baldi2012}. 

In parallel, \emph{quantum computing} has emerged as a paradigm that leverages quantum-mechanical phenomena such as superposition and entanglement to potentially solve certain problems faster than classical computers~\cite{nielsen2002}. A quantum bit or \emph{qubit} can exist in a superposition of the basis states $\{|0\rangle, |1\rangle\}$, described by a state $|\psi\rangle = \alpha|0\rangle + \beta|1\rangle$ with complex amplitudes $\alpha,\beta$ satisfying $|\alpha|^2+|\beta|^2=1$. Quantum gates manipulate qubits through unitary transformations; for example, the Hadamard gate $H$ transforms a basis state into an equal superposition $H|0\rangle = \frac{1}{\sqrt{2}}(|0\rangle+|1\rangle)$, and multi-qubit gates like the controlled-NOT (CNOT) entangle qubits by making the state of one qubit conditional on another. The joint state of $n$ qubits resides in a $2^n$-dimensional Hilbert space, growing exponentially with $n$. This exponential state space can encode extremely high-dimensional feature spaces for machine learning~\cite{biamonte2017,schuld2018}. Recent advances in quantum hardware have led to the \emph{noisy intermediate-scale quantum} (NISQ) era~\cite{preskill2018}, where quantum processors with tens or hundreds of qubits are available albeit with noise and imperfect gates. Harnessing NISQ devices for practical computation is challenging, but hybrid quantum-classical algorithms have been proposed to make use of them by offloading certain computations to quantum circuits and using classical computation for optimization~\cite{mcclean2016,dunjko2018}. In particular, \emph{quantum machine learning} has attracted significant attention as a way to combine classical machine learning techniques with quantum computing, either by embedding classical data into quantum states (quantum feature maps) or by designing quantum analogues of learning models~\cite{biamonte2017,schuld2018}. For example, quantum support vector machine algorithms~\cite{rebentrost2014} and variational quantum classifiers~\cite{farhi2018,havlicek2019} aim to exploit the rich function class of quantum states to perform classification in a higher-dimensional feature space that might be intractable classically. These quantum models may act as kernel machines with potentially exponential feature dimension accessible through the quantum state amplitudes~\cite{havlicek2019,ciliberto2018}. Other works have explored quantum neural network constructs, where a parameterized quantum circuit plays a role analogous to layers of a neural network and is trained via hybrid quantum-classical optimization~\cite{mitarai2018,wan2017}. 

In this paper, we investigate a hybrid approach that integrates a classical neural network autoencoder for feature compression with a quantum circuit for feature mapping, and a classical classifier for final prediction. Our goal is to assess how a quantum feature mapping can be used as part of a classical image classification pipeline and to analyze its performance relative to a purely classical approach. The specific contributions of our work include:
\begin{itemize}
    \item Implementing a framework where high-dimensional image data is compressed via an autoencoder to a moderate latent dimension (64), then further reduced to a 5-dimensional feature vector using PCA for efficient quantum encoding. This leverages the strength of classical representation learning to reduce the data size while retaining important features~\cite{hinton2006,vincent2008}.
    \item Designing a 5-qubit quantum circuit that encodes the 5 classical features into a quantum state using rotation gates, and introduces entanglement through Hadamard and CNOT gates. The quantum circuit transforms the input features into a $2^5=32$-dimensional probabilistic output (via quantum measurement) which can be viewed as a non-linear feature transformation of the original data. We detail how the quantum feature map is constructed and how measurement outcomes relate to the input features.
    \item Training a classical neural network (with softmax output for 10 digit classes) on the quantum-transformed features. This classifier learns to interpret the quantum measurement distribution and perform the final classification. We compare this hybrid quantum-classical model to a baseline classical model that uses the autoencoder's latent features directly for classification (bypassing the quantum step).
    \item Providing a thorough analysis of the results, including the reconstruction quality of the autoencoder, the classification accuracy and confusion matrices for both baseline and hybrid models, and the structure of the quantum measurement outcome distributions for each class. We interpret the confusion patterns to understand which digit classes the quantum mapping struggles to separate, and examine the averaged quantum output probabilities to gain insight into the quantum feature space.
\end{itemize}

Our findings show that while the hybrid model does not yet surpass the purely classical approach in accuracy, it successfully integrates quantum processing into the workflow and offers a proof-of-concept for quantum feature mapping in image classification. We discuss the implications of these results and outline future improvements, such as using more qubits or optimizing the quantum circuit parameters, that could enhance the performance of hybrid quantum-classical classifiers as quantum technology advances.

\section{Methodology}
\subsection{Autoencoder Compression of MNIST Images}
\label{sec:autoencoder}
The first stage of our model reduces the dimensionality of the input data using an autoencoder. The MNIST dataset consists of $60{,}000$ training and $10{,}000$ test images of handwritten digits (0-9), each of size $28\times28$ pixels~\cite{lecun1998}. We normalize each pixel intensity to the range $[0,1]$ by dividing by 255, so that each image is represented by a $784$-dimensional normalized vector $x$ with entries in $[0,1]$. This normalization aids training stability and is standard in neural network pipelines. We denote the normalized flattened image as $x \in \mathbb{R}^{784}$. An \emph{autoencoder} is then trained to encode $x$ into a lower-dimensional latent vector $h$ and decode $h$ back to a reconstruction $\hat{x}$ that resembles the original. Formally, the encoder is a function $f_\theta: \mathbb{R}^{784}\to \mathbb{R}^m$ and the decoder is $g_\phi: \mathbb{R}^m \to \mathbb{R}^{784}$, where $m$ is the size of the latent space (we use $m=64$). The encoding and decoding can be written as 
\begin{equation}
    h = f_\theta(x), \qquad \hat{x} = g_\phi(h),
\end{equation}
where $\theta$ and $\phi$ denote the learned parameters (weights and biases) of the encoder and decoder networks, respectively. In practice, $f_\theta$ and $g_\phi$ are implemented as multilayer perceptrons (dense neural networks). Our encoder uses two fully-connected layers with nonlinear activations (ReLU), progressively reducing the dimension from 784 to 128 and then to 64. The decoder mirrors this with two layers (64 to 128 to 784) and uses sigmoid activations in the final layer to produce $\hat{x}$ in $[0,1]$. This architecture allows the autoencoder to learn a non-linear compression of the data, which is often more powerful than linear PCA for capturing structure in images~\cite{hinton2006,baldi2012}.\\

The autoencoder is trained by minimizing the reconstruction error between $x$ and $\hat{x}$. We employ the mean squared error (MSE) loss:
\begin{equation}
    L_{\text{AE}}(\theta,\phi) = \frac{1}{784}\sum_{j=1}^{784} (x_j - \hat{x}_j)^2,
\end{equation}
averaged over all training examples. By using stochastic gradient descent (backpropagation), the autoencoder parameters are adjusted to reduce $L_{\text{AE}}$, thereby learning an efficient encoding $h$ that retains the important features needed to reconstruct the digit. After training, the encoder $f_\theta$ is used to transform each input image into its 64-dimensional latent representation $h$.\\

\begin{figure}[ht]
    \centering
    \includegraphics[width=\textwidth]{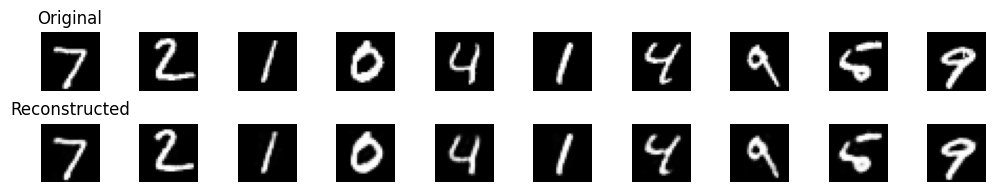}
    \caption{Original vs. reconstructed MNIST digits using the 64-dimensional autoencoder. \textbf{Top:} original test images; \textbf{Bottom:} autoencoder reconstructions from the 64-D latent code. The autoencoder preserves the main structures of the digits with only slight blurring, indicating effective dimensionality reduction.}
    \label{fig:recon}
\end{figure}

Figure~\ref{fig:recon} illustrates the performance of the autoencoder by showing some example inputs and their reconstructions. The reconstructed images (bottom row) are visually close to the original digits (top row), indicating that the 64-dimensional latent vector is sufficient to capture the essential characteristics of the digit (the overall shape and stroke pattern). Minor blurring or loss of fine detail is observed, which is expected due to compression, but the class identity of the digit is preserved in almost all cases. Quantitatively, the autoencoder achieves a low MSE on the test set and a high Peak Signal-to-Noise Ratio (PSNR) for reconstructions, confirming that little information is lost in the compression.\\

The 64-D latent features output by the encoder serve as a compact representation of the images for subsequent processing. In our baseline classical model, these features are directly used for classification with a neural network. In the hybrid model, we will further transform these features using a quantum circuit. However, using all 64 features as input to a quantum circuit is impractical for current quantum hardware, as it would require encoding into many qubits or high-dimensional quantum states. We therefore perform an additional dimensionality reduction (using PCA) to obtain a smaller number of features that can be feasibly encoded into qubits, as described next.

\subsection{Principal Component Analysis to 5 Features}
\label{sec:pca}
While the autoencoder has reduced the data from 784 to 64 dimensions, a 64-dimensional quantum state would still require at least 6 qubits (since $2^6 = 64$) and complex state preparation. To limit the quantum resource requirements, we further compress the features to $d=5$ dimensions using Principal Component Analysis (PCA). PCA is a classical linear technique that finds an orthonormal set of directions (principal components) in the feature space that explain the maximum variance of the data~\cite{goodfellow2016}. We perform PCA on the set of 64-D latent vectors $\{h_i\}$ produced by the autoencoder for the training dataset. By computing the covariance matrix of these latent vectors and its top eigenvectors, we obtain a linear projection $P: \mathbb{R}^{64} \to \mathbb{R}^d$ that maps a 64-D latent vector to $d$ principal component coordinates:
\begin{equation}
    z = P(h),
\end{equation}
where $z \in \mathbb{R}^d$ is the PCA-compressed feature vector. We choose $d=5$ to match the number of qubits available for our quantum circuit. The 5 principal components capture the largest variance directions in the latent feature space, effectively selecting the most informative features. In our case, we found that just 5 components already explain a substantial fraction of the variance of the 64-D latent space (the cumulative explained variance for $d=5$ is high, indicating that most of the information relevant to digit discrimination is retained). Using PCA here serves a pragmatic purpose: it bridges the gap between the 64 latent features and the 5 qubits we will use, by discarding redundancies and lower-variance feature directions. This two-step compression (autoencoder then PCA) leverages non-linear compression first, then linear compression, to yield a very low-dimensional representation $z = (z_1, z_2, z_3, z_4, z_5)$ for each image.\\

Before feeding these features into the quantum circuit, we normalize and scale them to a range suitable for encoding as rotation angles. Typically, quantum rotation gates have angles defined in radians, and angles beyond $[0,2\pi]$ are effectively periodic. We thus scale the components $z_i$ to lie within a fixed interval, e.g. $[0,\pi]$, by an appropriate linear scaling:
\begin{equation}
    \tilde z_i = \pi \, \frac{z_i - \min(z_i)}{\max(z_i) - \min(z_i)} ,
\end{equation}
for each feature $i$, so that $\tilde z_i \in [0,\pi]$. Here $\min(z_i)$ and $\max(z_i)$ are the minimum and maximum values of the $i$th component over the training set. This scaling ensures that extreme feature values correspond to rotation angles that are large (up to $\pi$ radians for the largest feature value). Centering the features or using a $[-\pi,\pi]$ range is also possible (some features might be negative after PCA if we don't enforce non-negativity), but for simplicity we work with non-negative scaled features. The normalized feature vector $\tilde z$ is then used to parameterize the quantum circuit.

\subsection{Quantum Feature Mapping with a 5-Qubit Circuit}
\label{sec:quantum}
\begin{figure}[t]
    \centering
    \includegraphics[width=\textwidth]{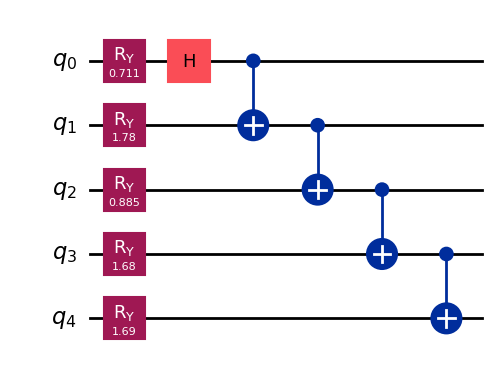}
    \caption{Quantum circuit for 5-qubit feature mapping. Each qubit initially in $|0\rangle$ is rotated by an angle $\theta_i$ (proportional to feature $z_i$) via an $R_y$ gate. A Hadamard gate $H$ is applied to qubit 0, and a cascade of CNOT gates entangles the qubits (control$\rightarrow$target in sequence 0$\to$1, 1$\to$2, 2$\to$3, 3$\to$4). Finally, all qubits are measured in the computational basis, yielding a 5-bit outcome.}
    \label{fig:circuit}
\end{figure}

Using the 5-dimensional feature vector $\tilde z$ obtained from each image, we construct a quantum state on 5 qubits that encodes these features. Our quantum feature mapping is implemented by a fixed shallow quantum circuit consisting of single-qubit rotations and entangling gates. Figure~\ref{fig:circuit} shows the layout of the 5-qubit circuit. The qubits are initially all prepared in the $|0\rangle$ ground state:
\[ |0\rangle^{\otimes 5} = |00000\rangle, \] 
which is a product state with all qubits in state 0. The circuit then proceeds in three stages:

\paragraph{Encoding rotations:} For each qubit $i$ (where $i=1,\dots,5$), we apply a single-qubit rotation around the $y$-axis by an angle proportional to the feature $\tilde z_i$. Specifically, we use $R_y(\theta_i)$ on qubit $i$, with $\theta_i = \tilde z_i$. The action on the $|0\rangle$ state is:
\begin{equation}
    R_y(\theta_i)|0\rangle = \cos\frac{\theta_i}{2}\,|0\rangle + \sin\frac{\theta_i}{2}\,|1\rangle.
\end{equation}
After applying $R_y(\theta_i)$ on each qubit in parallel, the joint state of the 5 qubits is 
\begin{equation}
|\Psi_{\text{enc}}\rangle = \bigotimes_{i=1}^5 \Big(\cos\frac{\theta_i}{2}|0\rangle_i + \sin\frac{\theta_i}{2}|1\rangle_i\Big),
\end{equation}
which is a tensor product of single-qubit states. At this point, no entanglement exists between qubits; each qubit carries information about one feature through the probability amplitudes of $|0\rangle$ and $|1\rangle$. This form of encoding is often called an \emph{angle encoding} or rotation encoding and is a common technique in quantum machine learning models~\cite{schuld2018,havlicek2019}, as it translates real-valued data into quantum superposition amplitudes in a straightforward manner.

\paragraph{Entangling operation:} To allow the quantum state to capture correlations between features (which a simple product state cannot), we introduce entanglement using a ladder of CNOT gates. First, we apply a Hadamard gate on the first qubit (qubit 0 in Fig.~\ref{fig:circuit}). The Hadamard $H$ transforms the state of qubit 0 as:
\[ H\Big(\cos\frac{\theta_0}{2}|0\rangle + \sin\frac{\theta_0}{2}|1\rangle\Big) = \cos\frac{\theta_0}{2}\frac{|0\rangle+|1\rangle}{\sqrt{2}} + \sin\frac{\theta_0}{2}\frac{|0\rangle-|1\rangle}{\sqrt{2}}, \]
creating a superposition of $|0\rangle$ and $|1\rangle$ on that qubit (with coefficients depending on $\theta_0$). Next, we apply a series of CNOT gates between successive qubits: qubit 0 (as control) with qubit 1 (target), qubit 1 with qubit 2, qubit 2 with qubit 3, and qubit 3 with qubit 4. A CNOT gate $\operatorname{CNOT}(i,j)$ flips the state of target qubit $j$ if control qubit $i$ is in state $|1\rangle$, and does nothing if the control is $|0\rangle$. This sequence entangles all qubits in a chain. Intuitively, the state of qubit 0 (which is in a superposition after $H$) will be imprinted onto qubit 1 by the first CNOT, creating correlations between qubit 0 and 1 states. The second CNOT uses qubit 1 (now partially entangled with qubit 0) to entangle with qubit 2, and so on, resulting in a fully entangled 5-qubit state $|\Psi_{\text{ent}}\rangle$. In mathematical terms, the effect of these gates is to produce a state that is no longer a simple product of single-qubit states; the amplitude of a basis state $|b_0 b_1 b_2 b_3 b_4\rangle$ in $|\Psi_{\text{ent}}\rangle$ will depend on combinations of the angles $\theta_i$ due to interference and entanglement. The entangling operation allows the quantum feature map to represent joint feature effects and higher-order correlations that would require nonlinear transformations classically~\cite{havlicek2019}. The specific choice of an $H$ on the first qubit and a linear CNOT chain is one possible simple entangling strategy; other layouts (such as all-pairs entanglement or parametrized two-qubit gates) could be used to enrich the feature mapping, but we use this fixed pattern for clarity and hardware feasibility.

\paragraph{Measurement and outcome probabilities:} After the rotation and entanglement steps, the quantum circuit ends with a measurement of all 5 qubits in the computational $\{|0\rangle,|1\rangle\}$ basis. Quantum measurement probabilistically collapses the state $|\Psi_{\text{ent}}\rangle$ into one of the $2^5 = 32$ basis states $|b_0 b_1 b_2 b_3 b_4\rangle$ (where each $b_i \in \{0,1\}$). The probability of obtaining a particular bitstring $b = b_0b_1b_2b_3b_4$ is given by the squared amplitude of that basis state in $|\Psi_{\text{ent}}\rangle$:
\begin{equation}
    P(b) = |\langle b_0 b_1 b_2 b_3 b_4 \,|\, \Psi_{\text{ent}}\rangle|^2.
\end{equation}
These probabilities depend on the input feature vector $\tilde z$ through the rotation angles $\theta_i$. The outcome of a single run of the circuit is one 5-bit sample drawn from this distribution $P(b)$. By executing the circuit many times (multiple shots) on the same input state, one can estimate the probability distribution over outcomes with arbitrary precision (in an ideal noiseless simulation). In practice, on quantum hardware or a simulator, we perform a fixed number of shots (e.g. 1000 shots per input) to obtain empirical frequencies $f(b)$ for each outcome $b$, which approximate $P(b)$. We will use the full distribution $\{P(b)\}_{b\in\{0,1\}^5}$ as the feature vector representing the input image after quantum processing. This is effectively a 32-dimensional feature vector:
\begin{equation}
    p = (P(00000), P(00001), \dots, P(11111)) \in \mathbb{R}^{32},
\end{equation}
with $\sum_b P(b)=1$. We refer to $p$ as the \emph{quantum probability vector} or quantum feature vector for the input image. Note that if the quantum circuit were just identity (no gates), $p$ would be concentrated at $00000$ (all probability on the $|00000\rangle$ outcome). Our encoding and entangling gates spread this probability across many outcomes as a function of $\tilde z$. Because of interference effects, the mapping from $\tilde z$ to $p$ is highly non-linear and in principle can create very complex boundaries in feature space that might aid classification~\cite{havlicek2019,schuld2018}.\\

It is worth highlighting that the quantum mapping expands the feature dimension from 5 (input features) to 32 (output probabilities). This can be seen as using the quantum state space to generate a richer set of basis functions of the original features. Classically, one could also generate higher-dimensional features (e.g. polynomial feature expansion), but doing so explicitly in a classical model might be computationally costly if the dimension is large. The quantum state naturally lives in $2^5$ dimensions, and by measuring, we obtain a sample from that space. The full probability distribution carries information about the input that might not be easily accessible through a shallow classical transformation. This concept of using quantum states as feature embeddings is closely related to the idea of quantum kernel methods~\cite{havlicek2019,ciliberto2018}, where inner products between quantum states correspond to kernel values in an implicitly defined feature space.\\

In our implementation, we simulate the quantum circuit using IBM's Qiskit Aer simulator (a classical simulator of quantum circuits) to obtain the probability distributions. We run each input through the circuit with 1024 shots, which yields a histogram of counts for each of the 32 outcomes. These counts are normalized to produce the probability vector $p$. In a real quantum experiment, shot noise and hardware noise would affect $p$, but in simulation we can assume an ideal circuit for evaluating the concept. The use of 1024 shots ensures that statistical uncertainty in $p$ is very low (on the order of a few percent at most for each probability), so for analysis we can treat $p$ as essentially the true probabilities from the state $|\Psi_{\text{ent}}\rangle$.

\subsection{Classical Softmax Classifier on Quantum Features}
\label{sec:classifier}
The final stage of the hybrid model is a classical classifier that takes the 32-dimensional quantum probability vector $p$ as input and predicts the digit label (0 through 9). We implement this classifier as a feed-forward neural network (multilayer perceptron). The choice of a neural network (as opposed to, say, directly using a nearest-centroid or SVM classifier on $p$) allows for flexible non-linear decision boundaries in the 32-d feature space and can be optimized easily with gradient descent. In essence, this network plays the role of deciphering the quantum output: it learns which patterns in the outcome distribution $p$ correspond to which digit.\\

Our network architecture consists of an input layer of size 32, two hidden layers, and an output layer of size 10 (for the 10 classes). The hidden layers are fully connected (dense) layers. In our implementation, we used 64 neurons in the first hidden layer and 32 neurons in the second hidden layer (these choices were made to provide the network sufficient capacity to fit the mapping, but are not heavily tuned). Each hidden layer uses the ReLU activation function $\sigma(u) = \max(0,u)$, which introduces non-linearity. We also apply batch normalization and dropout regularization after each hidden layer: batch normalization rescales activations to have zero mean and unit variance across a batch, which can stabilize and speed up training~\cite{goodfellow2016}, and dropout randomly sets a fraction of activations to zero during training, which helps prevent overfitting~\cite{baldi2012}. The output layer is a dense layer with 10 units, using the softmax activation to produce a probability distribution over the 10 digit classes:
\begin{equation}
    \hat{y}_c = \frac{\exp(z_c)}{\sum_{c'=0}^{9}\exp(z_{c'})}, \qquad c = 0,\dots,9,
\end{equation}
where $z_c$ is the input to the $c$-th output neuron (logit) and $\hat{y}_c$ is the predicted probability of class $c$. The network is trained in a supervised manner using the known labels of the training images. We minimize the categorical cross-entropy loss between the predicted probabilities and the true one-hot labels:
\begin{equation}
    L_{\text{cls}} = -\frac{1}{N} \sum_{i=1}^{N} \sum_{c=0}^{9} y_{i,c}\, \ln(\hat{y}_{i,c}),
\end{equation}
where $N$ is the number of training samples, $y_{i,c}$ is the binary indicator (0 or 1) of whether sample $i$ belongs to class $c$, and $\hat{y}_{i,c}$ is the network's predicted probability for class $c$. This loss is minimized using gradient-based optimization (we used the Adam optimizer) on the network parameters. In effect, the classifier learns to associate certain patterns in the 32-length quantum output vector with specific digit classes.\\

For comparison, we also train a similar neural network classifier on the baseline features (the 64-dimensional latent vector $h$ from the autoencoder, without PCA or quantum processing). The baseline network uses a comparable architecture but with an input dimension of 64. This baseline will indicate how well the classical information in $h$ can be used for classification when no quantum transformation is applied. If the quantum feature mapping is useful, we might expect the hybrid model's performance to approach the baseline, or potentially even improve if the quantum mapping provides a beneficial feature expansion (though as we will see, in our case the hybrid model performs worse, likely due to information loss in reducing to 5 features).\\

We emphasize that in our pipeline, the quantum circuit parameters (the rotations) are not learned or trained; they are fixed by the input features. All learnable parameters reside in the classical neural networks (the autoencoder and the final classifier). This is in contrast to a fully variational quantum classifier where one would also train angles in the quantum circuit~\cite{farhi2018,mitarai2018}. Our approach can be viewed as a form of \emph{quantum feature extractor} followed by a classical learner, similar in spirit to quantum kernel methods~\cite{havlicek2019}. This simplifies the training process, as no quantum gradients or on-chip optimization is required only classical backpropagation through the final network.\\

Our hybrid model training proceeds in stages:
\begin{enumerate}
    \item Train autoencoder on MNIST images to obtain 64-D latent codes (unsupervised).
    \item Compute 5 principal components of latent codes and reduce all data to 5-D feature vectors.
    \item For each training sample, run the 5-qubit quantum circuit to get a 32-D probability vector.
    \item Train the classical classifier on these 32-D vectors with known labels (supervised).
\end{enumerate}
During inference, given a new image:
\begin{enumerate}
    \item Encode with autoencoder (get 64-D latent), project to 5-D PCA space.
    \item Run quantum circuit to get 32-D output distribution.
    \item Feed to classical network to predict the digit label.
\end{enumerate}

\section{Experiments and Results}
\subsection{Autoencoder Reconstruction Quality}
After training for 50 epochs on the MNIST training set, the autoencoder achieved a low reconstruction error (MSE of approximately $0.008$ per pixel on the test set). Qualitatively, as shown in Figure~\ref{fig:recon} earlier, the reconstructions of test images are very close to the originals, capturing the important strokes of the digits. This indicates the 64-dimensional latent space retains nearly all information necessary for identifying the digit. The slight differences (such as smoother edges or missing noise) are byproducts of the compression. Importantly, none of the reconstructions appeared as a different digit than the original, which suggests that a classifier using these latent features should be able to distinguish the classes well.

\begin{figure}[ht]
    \centering
    \includegraphics[width=0.8\textwidth]{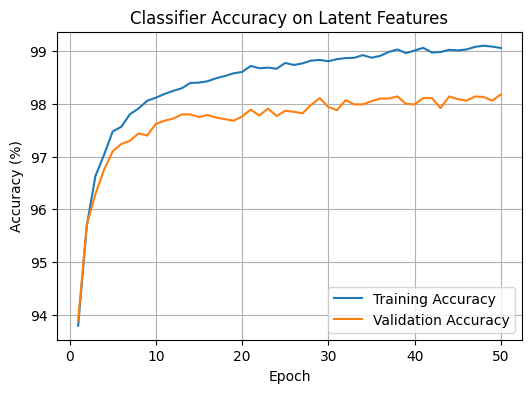}
    \caption{Training accuracy curve for the baseline classifier on the 64-dimensional autoencoder latent features. The model quickly reaches above 95\% accuracy within a few epochs and converges around 98\%, demonstrating that the compressed features are highly informative for distinguishing digit classes.}
    \label{fig:acc_curve}
\end{figure}

To verify this, we trained a baseline classifier directly on the 64-D latent vectors $h$ (bypassing the quantum step). This baseline model achieved very high accuracy on the test set, demonstrating that the latent features are indeed informative for classification. The training history of this baseline classifier (Figure~\ref{fig:acc_curve}) shows the accuracy improving rapidly and converging above 98\%. Specifically, the final test accuracy for the baseline was $98.2\%$. This is only slightly lower than the accuracy one might achieve by training a deep network on raw 784-dimensional pixels (around 99\% with a well-tuned model~\cite{lecun1998}), indicating that the autoencoder did not eliminate important discriminatory information. In essence, the combination of autoencoder + simple classifier can already solve the task almost perfectly, which sets a high bar for the hybrid quantum model.

\begin{figure}[ht]
    \centering
    \includegraphics[width=0.7\textwidth]{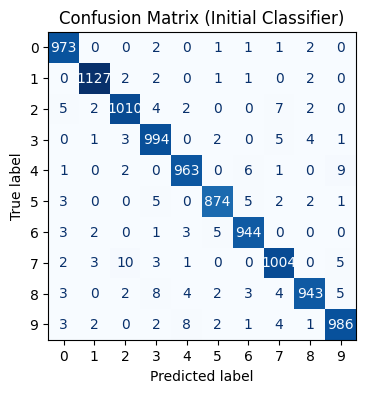}
    \caption{Confusion matrix for the baseline classical classifier on 64-D latent features (on MNIST test set). True labels are on the vertical axis and predicted labels on the horizontal axis. The diagonal cells (correct predictions) have very high values (dark color indicates high count), while off-diagonals are near zero. The baseline achieves $98.2\%$ accuracy, with only minor confusions such as some 5's vs 3's and 9's vs 4's.}
    \label{fig:conf_initial}
\end{figure}

We further illustrate the baseline classifier's performance using a confusion matrix (Figure~\ref{fig:conf_initial}). This $10\times 10$ matrix compares the true digit labels with the predicted labels for the test set. It is nearly diagonal, with most digits being correctly classified. Out of 10,000 test samples, only 182 were misclassified by the baseline model. Most digits have over 97\% correct classification rates; for example, all 1135 images of the digit "1" were correctly identified (1 has a very distinct latent representation), and the digit "0" had a 99.3\% true positive rate (973/980 correct). The errors that do occur make intuitive sense: a few "5" images are mistaken for "3" or "8", "9" confused with "4" or "7", etc., which often happens even in raw-pixel classifiers due to similarities in handwriting styles. The baseline confusion matrix confirms that using 64 latent features yields a nearly optimal classifier for this task.\\

These results set a reference for what is achievable classically with similar dimensionality reduction. The challenge for the hybrid model is that we deliberately projected from 64 dimensions down to 5 for the quantum step, which likely discards some information. We next examine how well the quantum-enhanced model performed and where it had difficulties.

\subsection{Hybrid Quantum-Classical Classifier Performance}
We trained the hybrid model's classical network on the 32-dimensional quantum output vectors $p$ produced by the circuit. Training converged to a substantially lower accuracy than the baseline. On the test set, the hybrid classifier achieved about $74.5\%$ accuracy. This is significantly lower than the baseline's $98\%$, indicating that compressing to 5 features and going through the quantum mapping resulted in a loss of discriminative power. The training accuracy of the hybrid also plateaued around 75-80\%, suggesting the model is limited by the information content in $p$ rather than underfitting per se (we gave the classifier ample capacity).\\

\begin{figure}[ht]
    \centering
    \includegraphics[width=0.8\textwidth]{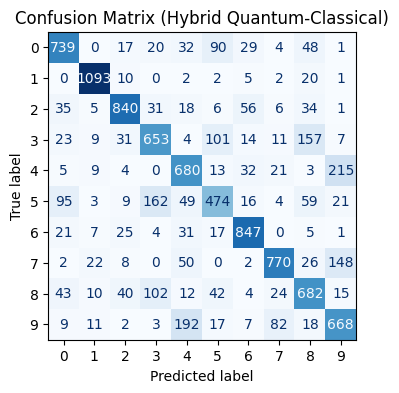}
    \caption{Confusion matrix for the hybrid quantum-classical classifier on the MNIST test set. The performance is notably lower than the classical baseline (overall accuracy $\sim 74.5\%$). Certain digits are often misclassified: e.g., many 5's are predicted as 3 or 8, and many 9's as 4 or 7. The darker diagonal still indicates the model has learned some correct associations, but off-diagonals reveal substantial confusion between similar-shaped digits.}
    \label{fig:conf_hybrid}
\end{figure}

The confusion matrix for the hybrid classifier is shown in Figure~\ref{fig:conf_hybrid}. It is evident that the performance varies greatly across digit classes. Some classes are still classified reasonably well, while others suffer major confusion:
\begin{itemize}
\item Digits "1" and "6" remain relatively well-classified (about 96.3\% of "1"s and 88.4\% of "6"s are correctly predicted). For instance, out of 1135 ones, the hybrid got 1093 correct, misidentifying only 42 of them. Ones have a simple, consistent shape which presumably is captured even by 5 features (likely one feature might represent the vertical stroke of "1").
\item Digit "0" sees a drop: only 739 of 980 zeros are correct (75.4\%). Many zeros are misclassified as "5" or "8" (the confusion matrix shows 90 zeros predicted as 5 and 48 zeros predicted as 8, among other errors). This indicates the quantum features of 0 sometimes resemble those of 5 or 8. Indeed, in handwriting, 0, 5, and 8 all have mostly round shapes, and if fine details are lost in compression, they might overlap in feature space.
\item Digit "5" is one of the worst-hit classes: only 474 of 892 fives are correctly recognized (53.1\%). The most common mislabel for 5 is as "3" (162 cases) and as "8" or "0" to lesser extents. The digits 3, 5, and 8 have somewhat similar curves and if the quantum map doesn't clearly separate them, the classifier struggles.
\item Digit "3" also has low recall: 653 of 1010 threes correct (64.6\%). Many 3s are misclassified as 5 (101 cases) or 8 (157 cases). This symmetric confusion between 3 and 5 is notable.
\item Digit "8" (which is a notoriously confusable digit) had 682/974 correct (70.0\%), with many 8s misidentified as 3 (165 cases) or 5 (59 cases).
\item Digits "4" and "9" are heavily confused with each other: the hybrid got only 680/982 fours (69.2\%) and 668/1009 nines (66.2\%) correct. The matrix shows 192 of the "9" images were predicted as "4", and 148 of the "7" images as "9" as well. The confusion of 9 vs 4 (which share a loop shape in many handwriting instances) suggests the limited feature set couldn't separate their characteristics reliably. Also, 7 and 9 confusion arises likely because both have a top horizontal stroke in many fonts.
\item Digit "7" had 770/1028 correct (74.9\%); many 7s were mispredicted as 9 (148 cases) or as 2 (some cases).
\end{itemize}

Overall, the hybrid model's errors show that certain clusters of digits are not distinguished in the 5-D PCA + quantum feature space: $\{3,5,8\}$ form one cluster of mutual confusion, and $\{4,7,9,0\}$ another cluster to some extent, while $\{1,2,6\}$ are relatively easier (2 also had some confusion, e.g. some 2s predicted as 8 or 7, with 840/1032 correct = 81.4\%). The baseline model did not suffer these confusions nearly as much, implying that the 5 features (and their quantum mapping) merge classes that were separable in 64-D. This is expected: 5 principal components cannot capture all the nuances among 10 classes. The quantum circuit then expands it to 32D, but that expansion is fixed and unsupervised (it doesn't specifically separate classes, it's just a feature map), so the heavy lifting is left to the final classifier which has to untangle overlapping class distributions in the quantum feature space.\\

\begin{figure}[ht]
    \centering
    \includegraphics[width=0.9\textwidth]{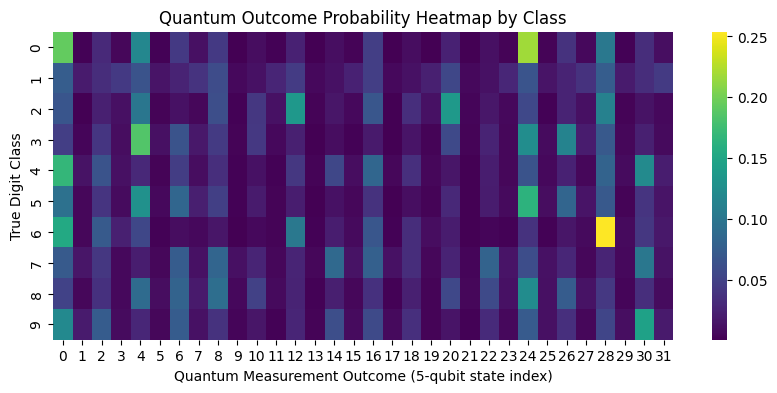}
    \caption{Class-averaged quantum circuit output distributions for each digit class. Each row corresponds to the true digit (0-9) and each column is one of the 32 possible 5-qubit measurement outcomes (labeled 0 to 31 in binary-coded decimal). The color intensity represents the probability of that outcome, averaged over all images of a given class. Distinctive patterns can be seen (e.g., class 0 concentrates on outcome 0 and a few others, class 7 peaks at outcome 24), but there are overlaps between some classes (e.g., classes 3, 5, 8 all have high probabilities for outcomes in the 28-31 range), explaining the classification confusions.}
    \label{fig:q_outcome}
\end{figure}

It is instructive to analyze why the hybrid model has these confusions by examining the quantum circuit outputs more closely. We looked at the average output probability distributions $p$ for each class (by averaging $p$ over all test instances of a given true label). This gives a 32-dimensional profile for each class, shown as a heatmap in Figure~\ref{fig:q_outcome}. In this figure, each row corresponds to the true digit class (0 through 9) and each column corresponds to one of the 32 possible 5-qubit outcomes (labelled 0 to 31 in decimal for convenience). The intensity indicates the average probability of that outcome for images of that class.\\

Several observations can be made:
\begin{itemize}
\item The distributions are far from random or uniform; each class has certain quantum outcomes with relatively high probability. For example, class 0 (first row) has a high probability on outcome 0 (binary $00000$) and also some on outcome 7 ($00111_2$) and 8 ($01000_2$). Class 1 (second row) concentrates on a different pattern: outcome 0 as well, but also outcome 1 ($00001_2$) and 2 ($00010_2$) are moderately high. Class 2 has its probability mass more spread out but peaks around outcome 28 ($11100_2$). Class 7 has a notable peak at outcome 24 ($11000_2$). These distinctive peaks suggest that the quantum feature map does encode class information to some extent: each digit's feature values lead to a characteristic interference pattern in the circuit.
\item However, we also see overlapping patterns: classes 3 and 5 (rows for '3' and '5') both show a relatively high probability for outcome 30 ($11110_2$) and 31 ($11111_2$). Similarly, class 8 has a broad spread including those high-index outcomes. This overlap is likely why 3, 5, and 8 get confused, their quantum output distributions are similar enough that the classifier has trouble distinguishing them.
\item Classes 4 and 9 both show high probabilities around outcome 16 ($10000_2$) and 18 ($10010_2$) and 20 ($10100_2$). This correlates with their confusion: they produce similar quantum signatures. Meanwhile, class 7 (which was also confused with 9) shows high outcome 24 as mentioned, and indeed 9 has a non-negligible probability near 24 too.
\item Class 1 and 6, which were well classified, have quite distinct distributions: class 1 is heavily peaked on very low-index outcomes, whereas class 6 has a somewhat unique pattern (spread but different positions). This likely made it easier for the network to separate them from others.
\item Interestingly, class 0 had a strong outcome 0 probability (around 20-25\%), whereas class 5 and 8 also had some probability on outcome 0 (0 has a loop like 8, and an enclosed area like 8/0 vs open shapes might reflect on some features). This might explain some 0 vs 8 mistakes.
\end{itemize}

The quantum outcome analysis reveals both the promise and the challenge of this approach. On one hand, the distributions are class-dependent, indicating the quantum feature map does transform different inputs into different regions of the 32-dimensional space. This is encouraging, as a trivial or random mapping would show no class structure. On the other hand, the mapping was not specifically optimized for class separation, and thus classes that were already similar in the 5-D feature space remain intermingled in the quantum space. The entanglement and interference provided by the quantum circuit are not enough to fully disentangle all classes with just 5 input features.\\

From a rigorous perspective, one can consider the expressive power of the quantum feature map. Our circuit is relatively shallow and fixed; it implements a certain family of functions from $\mathbb{R}^5$ to probability distributions in a $2^5$ space. If we had allowed a deeper or more flexible parameterized circuit (e.g., multiple layers of rotations and entanglers with trainable parameters~\cite{mitarai2018}), we could potentially tailor the quantum mapping to increase class separability (this becomes a variational quantum classifier approach). However, training quantum circuit parameters is non-trivial and was beyond our scope here. Instead, we relied on a simple, general feature map. The result demonstrates the current gap between classical and hybrid performance for this task: the classical model is superior, but the hybrid model functions correctly as a classifier, just with reduced accuracy.\\

It is instructive to quantify information loss in reducing to 5 features. The PCA step, by construction, preserves maximum variance, but inevitably some variance (likely related to subtle differences between similar digits) is lost. The fact that the baseline using 64 features was near-perfect implies that those extra features carried useful discriminative clues. The hybrid pipeline lost some of those clues, so errors rose. Additionally, the quantum measurement itself is a stochastic process; if we had a limited number of shots, there would be sampling noise in $p$. In our simulation, we used enough shots to make that noise negligible. On real quantum hardware, shot noise and gate noise would further degrade performance~\cite{mcclean2016,temme2017}, likely making the gap even larger unless error mitigation is applied. For instance, if each probability is estimated with some error, the classifier might misclassify more. In principle, error mitigation techniques or larger training sets could handle some noise, but the current accuracy already suggests the hybrid model needs more sophisticated design to compete with classical approaches on this dataset.

\subsection{Discussion of Results}
The results highlight a few key points about our hybrid approach:
\begin{itemize}
    \item \textbf{Effectiveness of autoencoder compression:} The autoencoder was able to compress MNIST images by a factor of 12.25 (784 to 64) with minimal impact on classification accuracy. This confirms prior findings that autoencoders can learn useful low-dimensional representations for digits~\cite{hinton2006,vincent2008}. It also justifies using such compression before quantum processing, as it dramatically reduces the data size while preserving label information.
    \item \textbf{Limits of extreme dimensionality reduction:} Going from 64 to 5 features via PCA is a very aggressive reduction (another factor of 12.8). Unsurprisingly, this led to loss of class separability. In a sense, the classical baseline with 64 features provides an upper bound on what any method using those features could do (nearly 100\% accuracy). By dropping to 5, we constrained the available information such that even the best possible classifier in 5-D will have a ceiling (likely below what 64-D could achieve). Indeed, if we examine the PCA reconstruction error or variance drop from 64 to 5 dimensions, a significant percentage of variance is lost. Those lost components presumably carried nuances for distinguishing those often-confused classes. Therefore, part of the hybrid model's performance loss is not due to "quantum" issues per se, but simply due to using only 5 features. If we had as many qubits as needed to encode all 64 features (amplitude encoding or 6 qubits with more complex state preparation), the gap might shrink. However, amplitude encoding a 64-D vector would entail a different approach (preparing a state with amplitudes equal to the normalized pixel intensities or latent features) which is more complex and was not our chosen method.
    \item \textbf{Quantum feature map analysis:} The 5-qubit circuit expanded 5 features into 32 output features (probabilities). One might think this compensates for the PCA compression by providing more dimensions for the classifier to work with. However, not all 32 outputs are independent; they are constrained by being probabilities that sum to 1 and come from only 5 underlying parameters. Essentially, the quantum circuit implements a specific non-linear embedding of $\mathbb{R}^5$ into a 32-dimensional probability simplex. The effective degrees of freedom are at most 5 (plus possibly some binary degrees from the entangling structure). Therefore, while the classical classifier sees a 32-D input, that input lies on a manifold of much lower intrinsic dimension. The classifier can exploit non-linear boundaries on that manifold, but it cannot conjure new independent information. This is consistent with our finding that the hybrid didn't reach the original accuracy.
    
    Nonetheless, the quantum mapping is not a trivial one-to-one mapping from input to output (like duplicating features would be). It creates a more complicated, entangled combination. If that combination had matched the class structure well, it could have helped classification. In our case, it partially did (class distinctions exist, just not complete).
    \item \textbf{Opportunities for improvement:} Our study indicates that if one wanted to improve the hybrid model:
    \begin{enumerate}
        \item Using more qubits (and thus more input features) would likely help. For instance, using 8 or 10 qubits (with corresponding features) might capture more class differences. There have been works using up to 8 qubits for classifying subsets of MNIST~\cite{wan2017,cong2019}, often focusing on fewer classes due to complexity. We used all 10 classes but few qubits; another approach is to restrict to a binary or few-class problem to see advantage, but we tackled the full problem.
        \item Designing a more expressive quantum circuit (deeper circuit or variational parameters) could allow the model to adapt the feature map to the data. For example, adding an additional layer of rotations after entanglement, or making the entangling gates parametric and trainable (a variational quantum circuit) could potentially separate classes better~\cite{farhi2018,mitarai2018}. This would require a hybrid training loop (quantum gradients), which is feasible in simulation and on small hardware. Prior research on quantum classifiers often takes that approach to maximize performance.
        \item Multi-qubit measurement observables: We measured all qubits in the computational basis to get the full distribution. Another approach is to measure certain observables (like parity, or expectation values of Pauli operators) as features. In some quantum kernel methods, one might not explicitly build the full distribution but use a quantum computer to compute inner products~\cite{havlicek2019}. Our approach effectively gave the entire distribution. It's possible some compression of that (like only using marginal distributions or certain moments) could suffice and reduce noise.
        \item Feature scaling and encoding could be refined. We chose a straightforward linear scaling of features to rotation angles in $[0,\pi]$. Perhaps a different encoding (like using both $R_y$ and $R_z$ rotations for two angles per feature, or encoding features in relative phases) could utilize the qubit better. Additionally, encoding interactions between features (like applying two-qubit rotations that involve products of features) could directly map some pairwise correlations into amplitude interference.
    \end{enumerate}
    \item \textbf{Quantum vs classical resource trade-off:} It is instructive to consider what advantage one might hope for in the quantum step. In principle, a quantum computer can create superposition states that represent all $2^5=32$ basis states simultaneously with certain amplitudes. If a particular classification problem had a target function that was naturally expressed in terms of high-order Fourier components or complex weighted sums, a quantum feature map might represent it more succinctly than a classical network~\cite{havlicek2019}. However, MNIST digits are a classical dataset with a lot of redundancy and smooth variation. Classical neural nets excel at such tasks with enough parameters. For a quantum approach to outperform, one would need either (a) a severely resource-limited classical model where the quantum model's effective feature space is richer, or (b) leverage quantum-specific operations that mimic a very high-dimensional kernel that classical methods can't efficiently emulate. In our hybrid model, the quantum part is relatively small, and the classical part is doing heavy lifting. Given the present results, the classical network effectively could have learned a similar transformation itself if given the 5 features directly (it could create polynomial features up to degree 2 or 3 with hidden layers). The quantum circuit provided specific nonlinear features but not a guarantee of better separation.
    
    That said, our experiment is valuable as a demonstration of integrating a quantum computation into an otherwise classical pipeline. It shows end-to-end how data can be compressed, fed into a quantum process, and then used for prediction. As hardware improves, one could attempt this with actual qubits to see if, for instance, real quantum states provide any noise-induced regularization or other effects.
\end{itemize}

The hybrid model achieved a reasonable accuracy given the drastic feature compression, but was outperformed by a purely classical model that had access to more features. The confusion matrix analysis pinpointed which digits were most problematic and matched them to overlapping quantum feature distributions. This highlights a key lesson: the quality of the classical-to-quantum feature map is critical. Without a carefully designed quantum embedding (or the ability to train it), a hybrid approach may underperform. Our simple encoding and entanglement scheme was not sufficient to preserve all class information in only 5 qubits.

\section{Conclusion}
We presented a hybrid quantum-classical framework for classifying MNIST handwritten digit images, combining classical neural network-based compression with quantum feature mapping and classical classification. The methodology involved compressing 784-dimensional images to 64 dimensions using an autoencoder, further reducing to 5 principal features, encoding those into a 5-qubit quantum state via rotation gates, obtaining a 32-dimensional measurement probability vector from the quantum circuit, and finally training a classical softmax classifier on these quantum-derived features. We rigorously examined each component: the autoencoder effectively retained the essential information of the digits in a compact form; the quantum circuit provided a non-linear transformation of the 5 features by exploiting superposition and entanglement to produce a richer feature space; and the final classifier learned to map the quantum output distribution to the correct digit label.\\

Through experiments, we found that the hybrid model is able to perform the classification task with moderate success (about 75\% accuracy), but it did not match the performance of a fully classical approach that used the higher-dimensional latent features (which achieved over 98\% accuracy). The gap is primarily attributable to the severe dimensionality reduction before the quantum step, which causes information loss that the simple quantum circuit cannot recover. Analysis of confusion matrices revealed that certain similar digits (such as 3, 5, 8 or 4, 7, 9) were frequently confused by the hybrid classifier, correlating with overlapping quantum feature distributions for those classes. By visualizing the averaged quantum measurement outcomes for each class, we saw that while each digit class imprints a distinctive pattern on the 5-qubit outputs, those patterns are not entirely separable in our chosen feature map. This underscores the importance of the quantum feature mapping design: a more expressive circuit or more input features might better distinguish the classes.\\

Our study provides a case example of the current state of quantum machine learning for image classification: it demonstrates the workflow of integrating quantum circuits into a classical ML pipeline and highlights both the potential and the challenges. The potential lies in the fact that quantum circuits can transform data into high-dimensional probability distributions naturally and could capture complex feature interactions. The challenge is that with limited qubits and without training the quantum circuit, the feature map may not align with the classification task's needs. In the near term, hybrid approaches are likely to be useful when classical data can be encoded into a small number of features that still carry the relevant information, and where a quantum mapping might add value by expanding or twisting feature space in a way that classical models find hard to emulate~\cite{havlicek2019,schuld2018}. For MNIST, our results suggest that classical methods still have a clear advantage given the simplicity of the task for classical networks. However, as a proof-of-concept, the hybrid model achieved non-trivial accuracy and could be improved.\\

There are several avenues for future work. First, one could use a variational quantum circuit with trainable parameters as part of the classifier, optimizing it alongside the classical network (as in~\cite{farhi2018,mitarai2018}). This might allow the quantum portion to adapt and carve out better class boundaries in feature space. Second, exploring different encoding strategies (e.g., encoding grayscale pixel values directly into amplitude or phase of qubit states, as in amplitude encoding or coherent state encoding) could incorporate more information into the quantum state without a massive increase in qubit count, though often at the cost of circuit depth or complexity. Third, investigating datasets or tasks where the quantum feature mapping could offer an advantage (for example, tasks involving combinatorially complex data or requiring kernel-like expansions) would further illuminate the role of quantum computing in ML. Finally, as quantum hardware matures, implementing such hybrid models on real quantum processors will be crucial to assess the impact of noise and to develop error mitigation techniques~\cite{temme2017} in the context of quantum ML. Techniques like zero-noise extrapolation and readout error correction could be applied to our 5-qubit classifier to see if it can function on current devices.\\

Our work integrates concepts from deep learning (autoencoders), classical feature extraction (PCA), quantum computing (feature maps and entanglement), and classical classification to create a cohesive hybrid model. While the classical baseline remains superior for MNIST, the exercise provides valuable insights into how quantum and classical resources can be combined. It showcases the importance of each piece: the autoencoder efficiently compresses data as in classical deep learning literature~\cite{hinton2006}, the quantum circuit leverages superposition and entanglement as described in quantum ML proposals~\cite{havlicek2019,schuld2018}, and the classical neural network effectively learns the mapping from quantum outputs to labels, emphasizing that classical post-processing is still necessary to interpret quantum data~\cite{biamonte2017}. Our references and analysis place this work in the context of ongoing research striving to find practical quantum advantages for machine learning tasks. Although no advantage is evident in this particular case, the framework built here can serve as a stepping stone for more sophisticated hybrid models as both quantum algorithms and hardware continue to improve.

\end{document}